\documentclass[a4paper,conference]{IEEEtran}
\ifCLASSINFOpdf
\else
\fi
\usepackage[pdftex]{graphicx}
\usepackage{cite}
\usepackage{flushend}

\newcommand{\wfig}[1]{Fig.\ref{fig:#1}}
\newcommand{\wfigure}[1]{Figure \ref{fig:#1}}

\newcommand{\wtable}[1]{Table \ref{tab:#1}}
\newcommand{\wsec}[1]{\textbf{\ref{sec:#1}}}

\begin{document}
%
\title{Bitstream-Based JPEG Image Encryption \\with File-Size Preserving}

\author{\IEEEauthorblockN{Hiroyuki KOBAYASHI}
\IEEEauthorblockA{
 Tokyo Metropolitan College \\
 of Industrial Technology,\\
 Tokyo, JAPAN\\
 Email: hkob@metro-cit.ac.jp}
\and
\IEEEauthorblockN{Hitoshi KIYA}
\IEEEauthorblockA{
 Tokyo Metropolitan University\\
 Tokyo, JAPAN\\
Email: kiya@tmu.ac.jp}
}


%


\maketitle



%
\IEEEpeerreviewmaketitle

\begin{abstract}
An encryption scheme of JPEG images in the bitstream domain is proposed.
The proposed scheme preserves the JPEG format even after encrypting the images, and the file size of encrypted images is the exact same as that of the original JPEG images.
Several methods for encrypting JPEG images in the bitstream domain have been proposed.
However, since some marker codes are generated or lost in the encryption process, the file size of JPEG bitstreams is generally changed due to the encryption operations.
The proposed method inputs JPEG bitstreams and selectively encrypts the additional bit components of the Huffman code in the bitstreams.
This feature allows us to have encrypted images with the same data size as that recoded in the image transmission process, when JPEG images are replaced with the encrypted ones by the hooking, so that the image transmission are successfully carried out after the hooking.
\end{abstract}

\begin{IEEEkeywords}
JPEG, Encryption, File-size preserving, Bitstream-based
\end{IEEEkeywords}

\section{Introduction}

%
%

Due to the spread of digital cameras and smart phones, opportunities to use digital images are increasing.
Generally, captured images are immediately JPEG encoded and stored.
These images are not only stored on personal devices, but also are often uploaded to cloud providers, such as social networks, cloud photo storage services, and so on.
Most the cloud providers accept only limited file formats like JPEG.
In addition, such cloud environments are based on the reliability of the providers, but they are not a reliable situation in terms of privacy preserving for users. 
Therefore, various image encrypting methods have been studied for compressed image data.

Encryption then Compression(EtC) systems \cite{AEtCSfJMJS,APSIPA3,APSIPA15,APSIPA16,APSIPA17,APSIPA18,APSIPA19,APSIPA24,APSIPA25,APSIPA2017Chuuman,2018-1}
are methods to encrypt for images before encoding.
Some of these methods have the compatibility with international compression standards, and enable privacy preserving decompression and compression.
The compression performances of EtC systems are almost as same as one of the original images, but the file sizes are slightly different from those of compressed images without encryption.

Several bitstream based encryption methods have also been proposed\cite{ESwGMCfJ2I,Imaizumi1,PSofJ2IwGMC,LpBsbJE,JEwFSP,EJIRUBwFC}.
For JPEG 2000 images, some bitstream-based encryption methods have been proposed in consideration of the generation of special marker codes and without changing the file size
\cite{ESwGMCfJ2I,Imaizumi1,PSofJ2IwGMC}.
Even for JPEG images, some bitstream-based encryption methods have been proposed, but the file size of the encrypted bitstream has changed by the occurrences or disappearances of the JPEG marker code \cite{LpBsbJE,JEwFSP,EJIRUBwFC}.
For example, in \cite{LpBsbJE}, encryption is performed while keeping the file size by rearranging the run lengths of AC coefficients.
However, occurrences or disappearances of the pseudo marker code have not been studied.
In this case, the encrypted bitstream can not be correctly decoded or the file size of the encrypted bitstream is changed.
On the other hand, in the methods of \cite{JEwFSP,EJIRUBwFC}, bitstream-based block scrambling and coefficient scrambling are implemented.
In these methods, in order to accurately hold the JPEG format, encryption processing is executed the byte stuffing operation which prevent accidental generation of markers by the arithmetic encoding procedures.
As a result, It has been shown that the file size changes by several bytes before and after encryption.

In this paper, we propose a bitstream-based JPEG encryption scheme that makes the file size exactly equal to the original.
The proposed method guarantees a constant file size by providing a mechanism to avoid occurrences / disappearances of the JPEG marker code.
This feature allows us to have encrypted images with the same data size as that stored in the image transmission process, when JPEG images are replaced with the encrypted ones by the hooking, so that the image transmission are successfully carried out after the hooking.

\section{JPEG Bitstream and Its Byte Stuffing}
\label{sec:fileSizeChange}
\subsection{JPEG Bitstream}
\begin{figure}[tbp]
\centering
\includegraphics[scale=0.4]{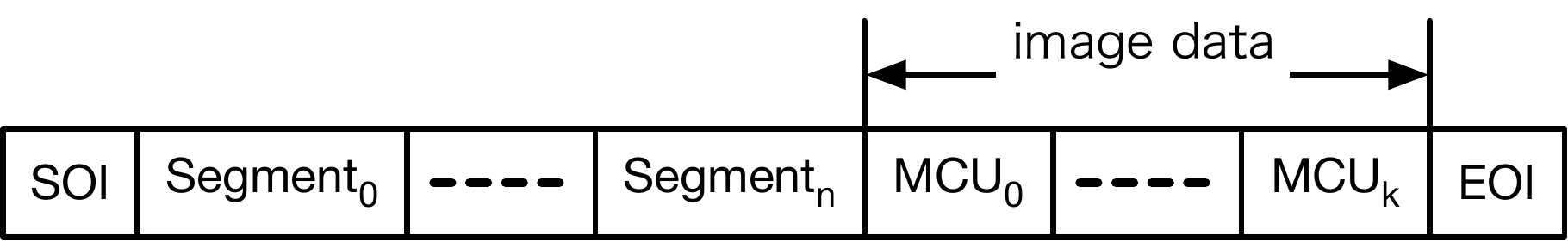}\\
(a) JPEG bitstream\\[2mm]
\includegraphics[scale=0.4]{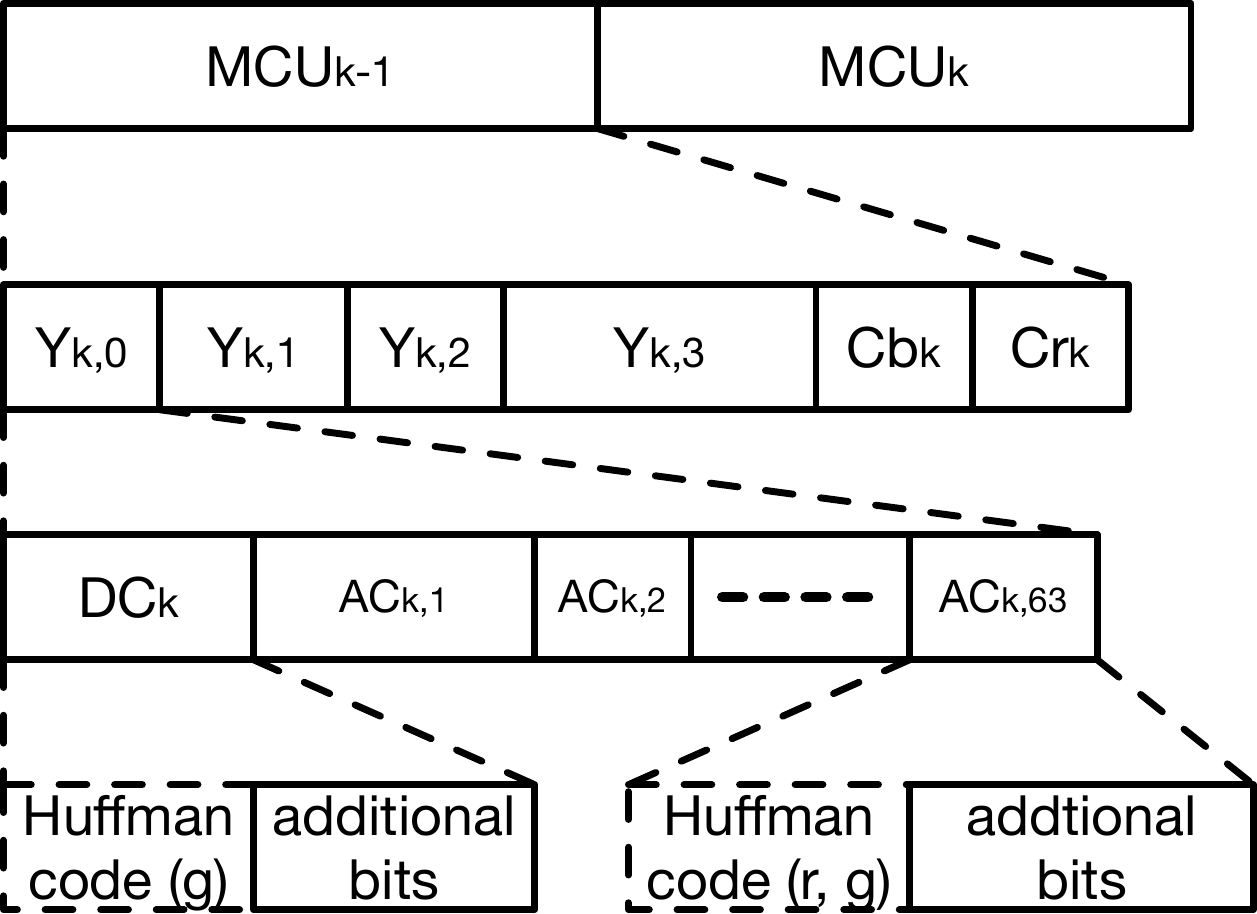}\\
(b) MCU\\
\includegraphics[scale=0.4]{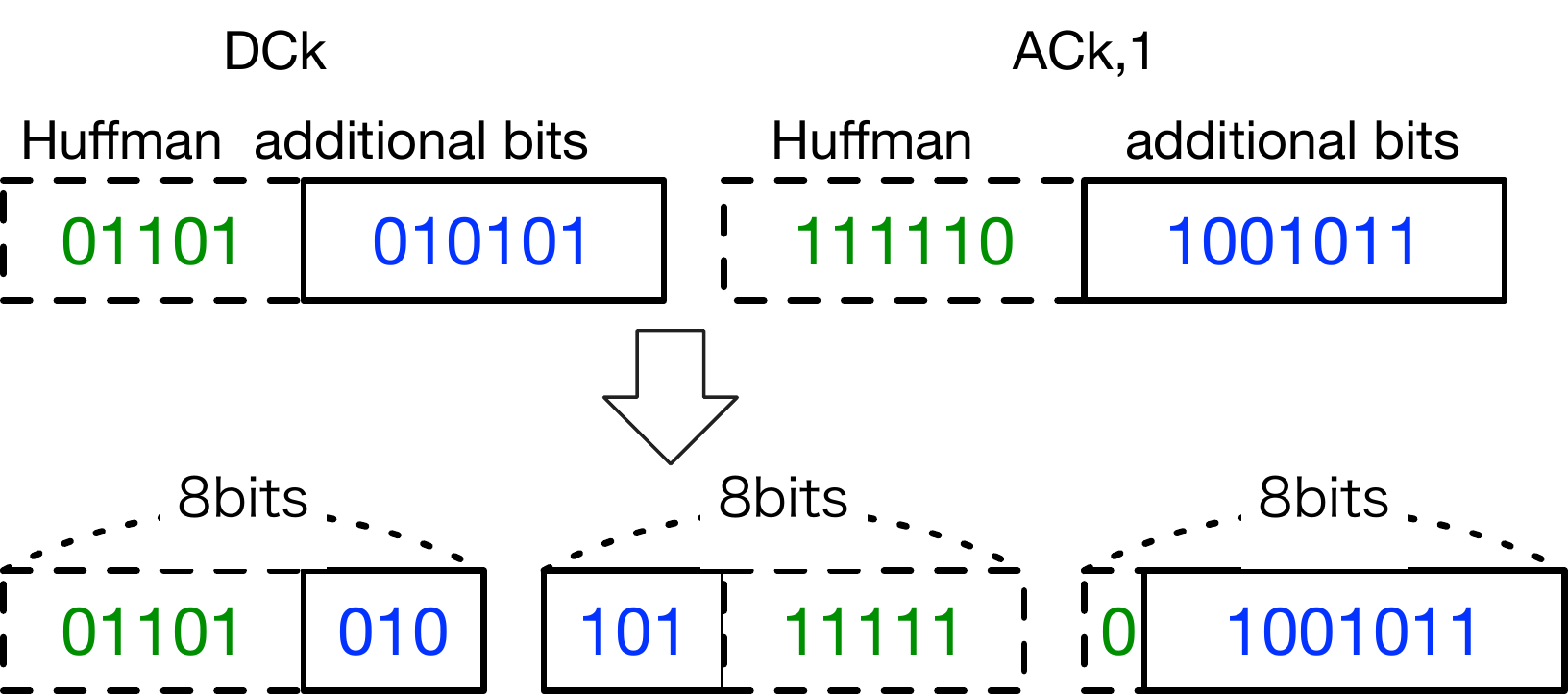}\\
(c) Representation with byte-aligned code
\caption{Structure of a JPEG bitstream}
\label{fig:bitstream}
\end{figure}

\wfigure{bitstream}(a) shows the structure of a JPEG bitstream.
SOI and EOI are the marker codes which correspond to ``Start of Image'' and ``End of Image'', respectively.
JPEG bitstreams have some marker segments (``Segment'' in \wfig{bitstream}(a)) which store information to be used for data decoding such as quantization tables, Huffman tables, and so on.
Each marker segment starts with a marker code.
The marker codes are special two-byte codes where the first byte is ``FF'' and  the second byte is a value between ``01'' and ``FE''.

\wfigure{bitstream}(b) shows the structure of the ``image data'' in \wfig{bitstream}(a).
``Image data'' consists of multiple MCUs (Minimum Coded Unit).
\wfigure{bitstream}(b) is an example in the case of 4:2:0 color subsampling.
Therefore, each MCU has four Y(luminance) blocks, one subsampled Cb block, and one subsampled Cr block.
Moreover, each block has $DC_k$ which is the difference value from the DC coefficient of previous block, and 63 AC coefficients $AC_{k,n}$ ($n=1\cdots63$).
Each coefficient has a Huffman code part corresponding to the group number ($g$) that determines the range of values, and an additional bits part for uniquely identifying values within the range.
In the case of AC coefficients, the Huffman code also includes the run length ($r$) of the zero value until a significant value exists.

\wfigure{bitstream} (c) is an example of Huffman code and additional bits.
Since each data is composed of variable length bits, they are stored in byte units.

\subsection{Difficulty of file size preserving}
\begin{figure}[tb]
\includegraphics[scale=0.4]{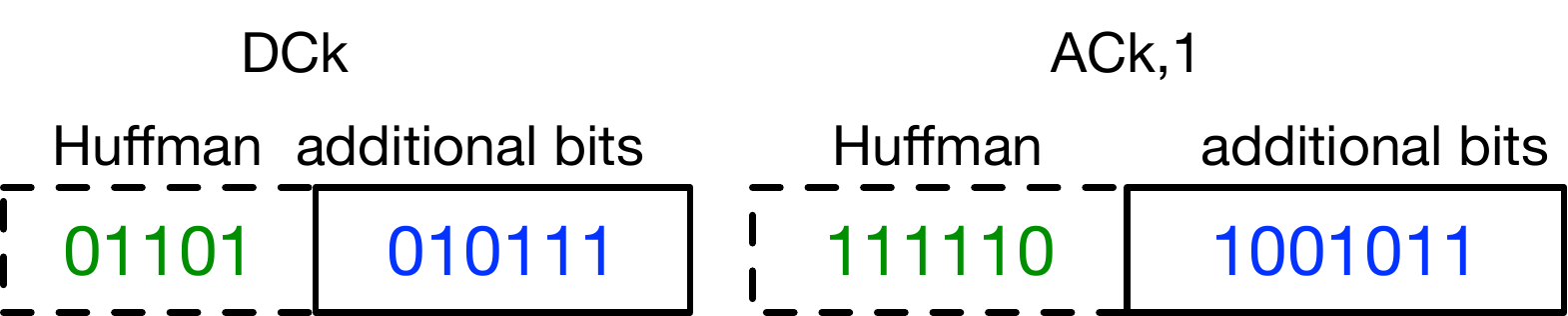}\hfill\null\\
\hspace{5em}(a) Entropy coded DCT coeffcients\\
\includegraphics[scale=0.4]{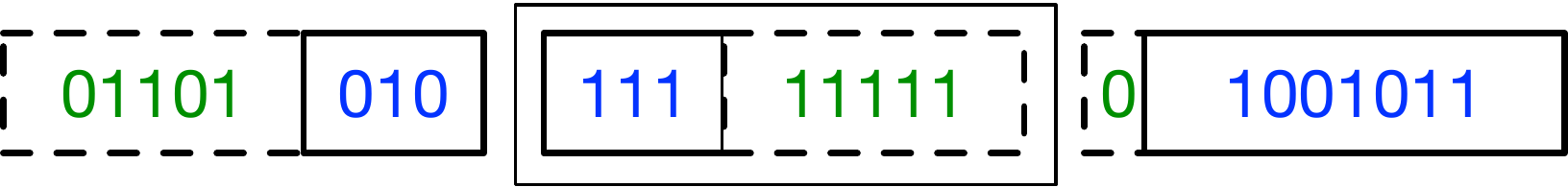}\hfill\null\\
\hspace{5em}(b) Byte-based packing\\
\includegraphics[scale=0.4]{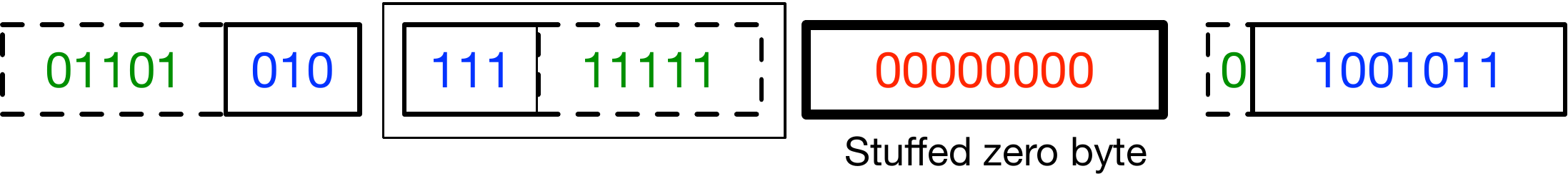}\\
\hspace{5em}(c) Byte stuffing
\caption{Byte stuffing in entropy-coded data segment}
\label{fig:ff00}
\end{figure}

\wfigure{ff00} illustrates an example of entropy-coded data segment, where $DC_k$ in \wfig{ff00}(a) is entropy-coded DC coefficient data and $AC_{k, 1}$ is the entropy-coded data of the first AC component in the $k$-th block, respectively.
Each data have Huffman code part and additional bits part.
To generate the JPEG bitstream, byte-based packing is first applied to the entropy-coded data in \wfig{ff00}(a), as shown in \wfig{ff00}(b).
The byte ``FF'', i.e. ``11111111'', which corresponds to a marker code, may be produced due to the byte-based packing.
Therefore, finally, in order to ensure that the marker does not occur within an entropy-coded segment, any ``FF'' byte in either a Huffman or additional bits, is followed by a ``stuffed'' zero byte, whose operation is called as `byte stuffing', as shown in \wfig{ff00}(c)\cite{JPEG}.

Note that the file size of the stream (b), is not the same as that of the stream (c).
We have to consider the byte stuffing operation to preserve the same file size when JPEG images are encrypted.


\section{Proposed Structure}

We propose a new bitstream-based JPEG image encryption method which allows us to exactly preserve the same file size as the original JPEG bitstream.

\subsection{Outline of the proposed structure}

\begin{figure}[tbp]
 \centering
 \includegraphics[width=3.3in]{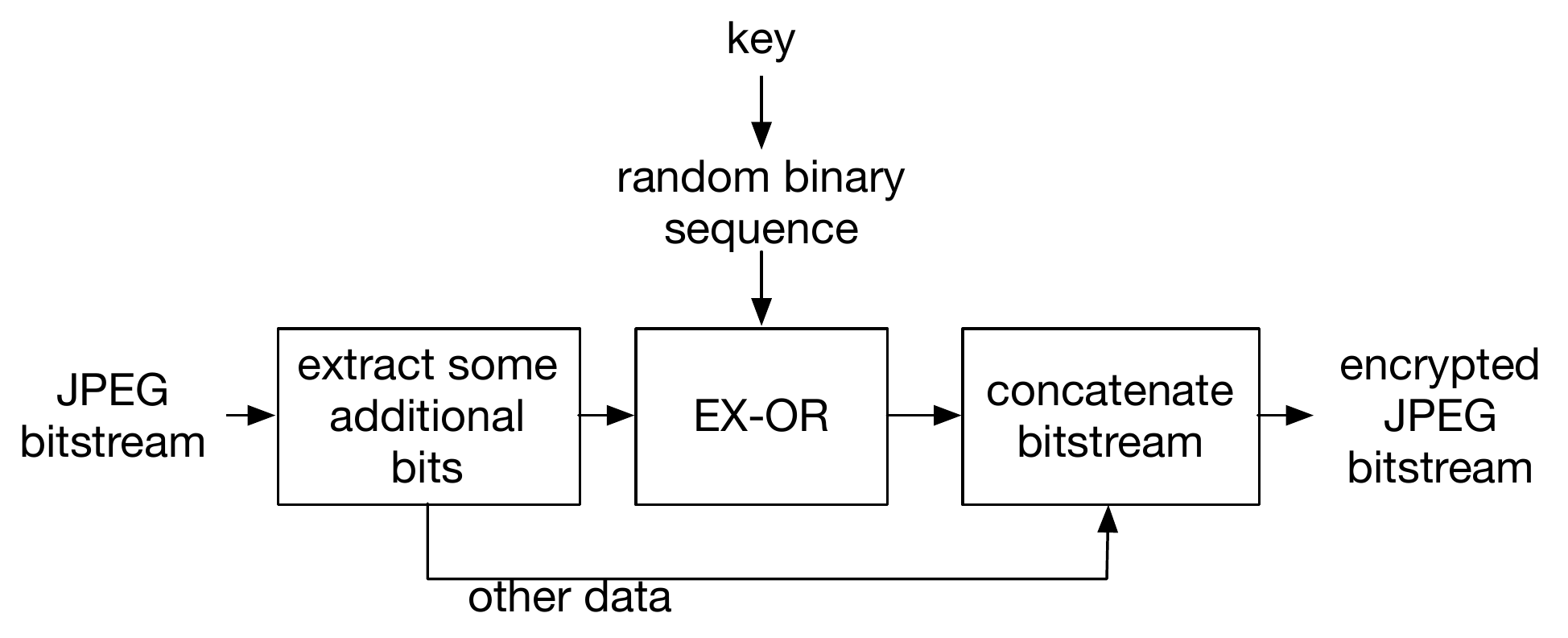}\\[-4mm]
 \caption{Outline of the proposed method}
 \label{fig:proposed}
\end{figure}

Bitstreams encrypted by the proposed method have not only the same file sizes, but also the compatibility with JPEG decoders.
Some of only additional bits fields that satisfy conditions are encrypted to keep the compatibility with JPEG decoders.
\wfigure{proposed} illustrates the outline of the proposed method.
The procedure of the proposed method is summarized as follows.
\begin{enumerate}
\item Analysis, byte-by-byte, the entropy-coded data segment and extract additional bits from a byte that satisfies two conditions: the byte includes both Huffman code and additional bits, and the Huffman code includes at least one ``0'' bit.  
\item Generate a random binary sequence with a secret key.
\item Carry out exclusive-or operation between only extracted additional bits and the random sequence generated in 2), and replace the additional bits with the result.
\item Produce an encrypted bitstream by combining the encrypted additional bits with other data without any encryption.
\end{enumerate}

\subsection{Encryption considering occurrences or disappearances of `FF00'}
\label{sec:condition}
If all additional bits are simply encrypted, there is a possibility to generate or lose ``FF''.
Therefore, in the proposed method, only limited additional bits are encrypted based on exclusive-or operation with a random binary sequence.

\begin{figure}[tb]
\centering
\includegraphics[width=8cm]{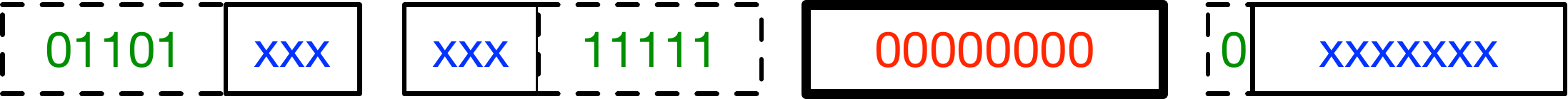}\\
\caption{Example for determination whether encrypt or not}
\label{fig:check}
\end{figure}

In \wfig{check}, the additional bits  in \wfig{ff00}(c) were replaced with 'x'.
Using \wfig{check}, we describe an outline of determination whether encryption is possible or not.
\subsubsection{The first byte}
The data consist of a 5-bit Huffman code and 3-bit additional bits.
Even if all the additional bits are 1, the entire byte never becomes ``FF''.
Therefore, the additional bits part is able to be encrypted.
\subsubsection{The second byte}
The data consist of 3-bit additional bits and a 5-bit Huffman code.
In the original data, since the additional bit was ``111'' and the remaining Huffman code was `111111'', ``FF'' was composed as the whole byte.
If any bit of the additional bit is changed to 0 by encryption, the entire byte is not `FF' and `00' of the third byte is not inserted.
Since this causes a file size change, the additional bits of the second byte are not encrypted.
\subsubsection{The third byte}
The data are padding data because the second byte is ``FF''.
Since this is not an additional bit, it is not encrypted.
\subsubsection{The last byte}
The data consist of a 1-bit Huffman code and 7-bit additional bits.
Even if all the additional bits are 1, the entire byte never becomes ``FF''.
Therefore, the additional bits part is able to be encrypted.

By analyzing the Huffman code in the byte as described above, it is possible to determine whether encryption is possible or not.
In summary, the following bytes are not encrypted.
\begin{enumerate}
\item The whole 8 bits are Huffman codes.
\item The whole 8 bits are additional bits.
\item ``00'' byte immediately after ``FF''.
\item Huffman code and additional bits are included and the all bits of Huffman code are `1'.
\end{enumerate}


\section{Experimental Results and Discussion}

Some simulations were carried out to demonstrate the effectiveness of the proposed method.
For the simulations, we used the reference software distributed by JPEG\cite{reference} with 4:2:0 chroma subsampling.

\subsection{Image quality evaluation of encrypted image}
\begin{figure}[tb]
\centering
\begin{tabular}{cc}
\includegraphics[width=4cm]{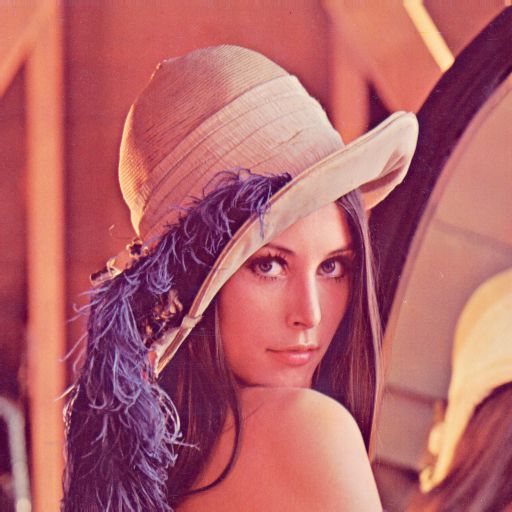}
&
\includegraphics[width=4cm]{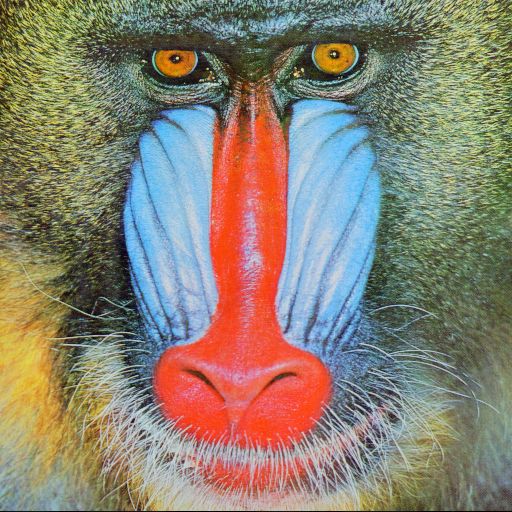}
\\
(a) Original ($Q=80$) & (b)Original ($Q=80$)\\
(lena) & (mandrill) \\
\includegraphics[width=4cm]{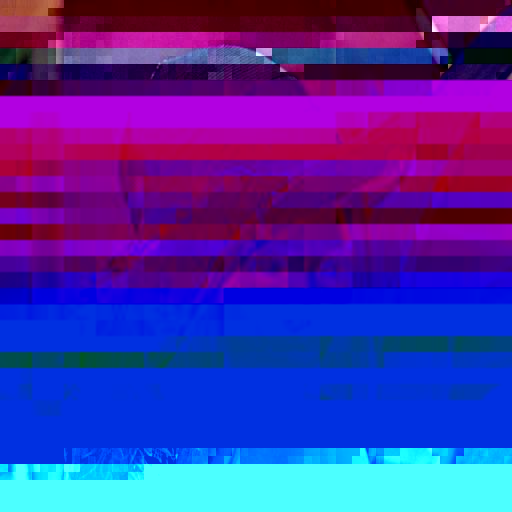}
&
\includegraphics[width=4cm]{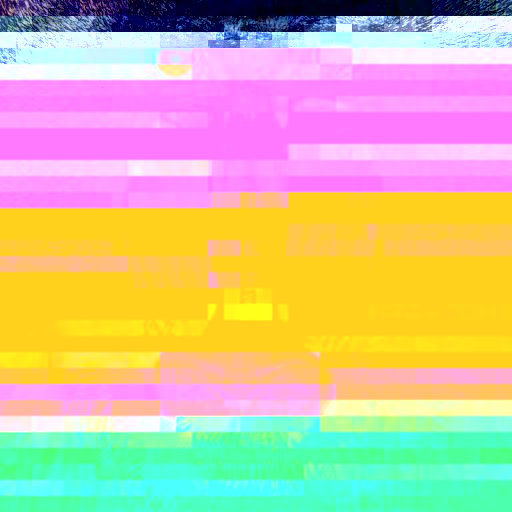}\\
(c) DC component only &
(d) DC component only \\
(lena) & (mandrill)\\
\includegraphics[width=4cm]{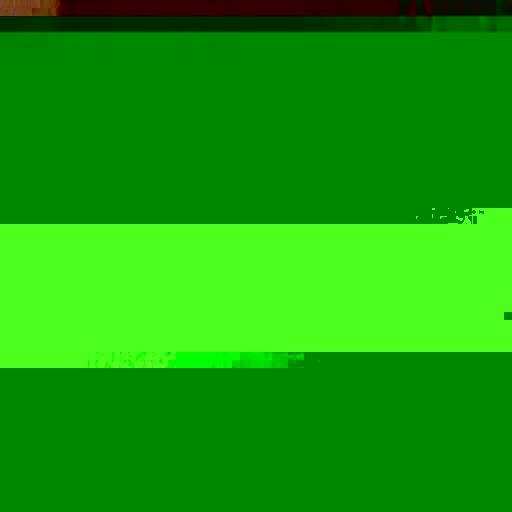}
&
\includegraphics[width=4cm]{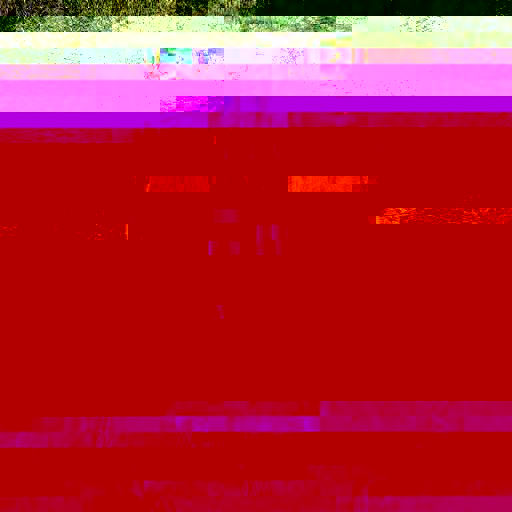}\\
(e) Both Components &
(f) Both Components \\
(lena) & (mandrill)
\end{tabular}
\caption{Original and encrypted results (lena, mandrill)}
\label{fig:proposedResult}
\end{figure}

First, the image quality of encrypted images was evaluated.
\wfigure{proposedResult}(a) and (b) were standard images, ``lena'' and ``mandrill'' encoded by JPEG with Q-factor 80, respectively.
\wfigure{proposedResult}(c) to (f) were decoded images with a standard JPEG decoder from the encrypted images by the proposed scheme.
In \wfig{proposedResult}(c) and (d), the additional bits in only DC components were encrypted.
On the other hand, in \wfig{proposedResult}(e) and (f), the additional bits in both DC and AC components were encrypted.

The encrypted image in \wfig{proposedResult}(b) has slightly visible information on the original one, because the AC components were the same as the original ones.
On the other hand, the encrypted image in \wfig{proposedResult}(c) had less visible information than \wfig{proposedResult}(b).



\subsection{The number of bytes to be encrypted}
\wtable{count} indicates the ratio of encryption applied to additional bits in ``image data''.
Data excluded from encryption are data which satisfy conditions 2) and 4) of \wsec{condition}.
Due to the increase of Q-factors, the proportion of encryption targets decreased.
This is because the number of bytes corresponding to the condition 2) necessity  increases due to the increase of Q-factors.

\begin{table}[tb]
\centering
\caption{The number of bytes to encrypt / bytes to be excluded from encryption
(Lena image)}
\label{tab:count}
(a) $Q=50$\\
\begin{tabular}{c||c|c|c}
target & \shortstack{\# of bytes \\excluded from \\encryption[byte]} & \shortstack{\# of encrypted\\ bytes[byte]} & \shortstack{Percentage of\\encrypted bytes[\%]}\\ \hline
DC only & 70 & 4,729 & 98.5 \\
AC only & 172 & 9,591 & 98.2 \\
Both& 197 & 13,467 & 98.6 \\
\end{tabular}\\
(b) $Q=80$\\
\begin{tabular}{c||c|c|c}
target & \shortstack{\# of bytes \\excluded from \\encryption[byte]} & \shortstack{\# of encrypted\\ bytes[byte]} & \shortstack{Percentage of\\encrypted bytes[\%]}\\ \hline
DC only & 420 & 6,306 & 93.8 \\
AC only & 460 & 20,931 & 97.8 \\
Both  & 741 & 26,104 & 97.2 \\
\end{tabular}\\
(c) $Q=95$\\
\begin{tabular}{c||c|c|c}
target & \shortstack{\# of bytes \\excluded from \\encryption[byte]} & \shortstack{\# of encrypted \\bytes[byte]} & \shortstack{Percentage of\\encrypted bytes[\%]}\\ \hline
DC only & 1,758 & 7,152 & 80.2 \\
AC only & 3,225 & 59,063 & 94.8 \\
Both & 4,336 & 65,274 & 93.8 \\
\end{tabular}
\end{table}

\subsection{File-size preserving}

Next, we compare our method with the previous works\cite{EJIRUBwFC,AEtCSfJMJS}, in terms of the file sizes.
\wtable{changeSize} shows the file sizes of encrypted JPEG images under various conditions.
From this table, JPEG images encrypted by the proposed method had exactly the same file sizes as those of the original ones.
However, other encryption methods could not preserve the same file sizes, because they do not consider the effect of byte stuffing.

%

\begin{table}[tb]
\centering
\caption{Length of original and encrypted images (difference)[byte], for lena image}
\label{tab:changeSize}
\begin{tabular}{c|c|c|c}
Q-factor & 50 & 80 & 95 \\ \hline
Original & 24,279 & 43,879 & 106,548 \\ \hline
\textbf{Proposed} & \textbf{24,279(0)} & \textbf{43,879(0)} & \textbf{106,548(0)} \\ \hline
Cheng\cite{EJIRUBwFC} & 24,281(+2) & 43,865(-14) & 106,553(+5) \\ \hline
EtC\cite{AEtCSfJMJS} & 24,767(+488) & 44,487(+608) & 108,262(+1,714) \\ \hline
\end{tabular}
\end{table}

\section{Conclusion}
In this paper, we have proposed an encryption method that allows us to preserve the same file size before and after the encryption.
In the encryption process, the proposed method considers the effect of byte stuffing and guarantees a constant file size by providing a mechanism to avoid the occurrence or disappearance of the JPEG marker code.
By preserving the file size, it can be expected that the image transmission are successfully carried out after the hooking encryption process.

\end{document}